# The journey from Planck distribution to Bose statistics: From classical to quantum mechanics and beyond

SHREETAM DASH [1] and PRASANTA K. PANIGRAHI [1,2]

[1] *Center for quantum science and technology Siksha 'O' Anusandhan Deemed to be University, Khandagiri Square, Bhubaneswar 751030, Odisha, India*

[2] *Indian Institute of Science Education and Research Kolkata, Mohanpur, 741246 West Bengal, India*

**Abstract:** In 1924, Satyendra Nath Bose's pioneering work laid the foundation for Bose-Einstein statistics, that describes particles with integral spins. His derivation of Planck's law for blackbody radiation bypassed classical assumptions, relying instead on the statistical mechanics of light quanta. Bose's methodology addressed limitations in existing theories, such as Einstein's dependence on classical concepts like Wien's displacement law and Bohr's correspondence principle. Further, his work underscored the incompatibility between classical electrodynamics and quantum theory, proposing innovative statistical approaches to thermodynamic equilibrium. The insights from Bose's work extend beyond theoretical physics. As was immediately noticed by Einstein, for temperatures below a critical threshold, Bose-Einstein statistics predicts the formation of a Bose-Einstein condensate (BEC), where particles condense en-masse into the ground state. This quantum phenomenon on a macroscopic scale opened avenues to explore new technologies in recent times, apart from throwing light on new phases of matter. This article revisits Bose's ground-breaking contributions, highlighting their enduring impact on quantum mechanics, statistical physics and field theory.

**Keywords:** Bose Statistics, Light Quantum Hypothesis, Bose-Einstein Condensate

# 1. Introduction

This year, 2024 marks the hundredth year of Bose statistics, ushered in by a remarkable four-page derivation of Planck's distribution law by Satyendra Nath Bose, purely through counting of occupation numbers of the light quanta (photons) in the phase space cells. The name 'photon' for light quantum was coined much later by the remarkable chemist G.N. Lewis in the year 1926.

The pinnacle of classical mechanics in the 19$^{th}$ century, when Newton's laws of particle dynamics and Maxwell's laws for describing the movement of light waves led to an almost complete understanding of experimental observations, both at the laboratory and cosmic scales. The problem with classical laws started when it failed to explain the intensity distribution of thermal radiation emitted from a blackbody in the form of a hot sphere with the light coming out from a tiny hole . The low and high-temperature ends of this characteristic radiation could be explained by Wein's and Rayleigh-Jean's approaches respectively, that were unable to match the complete distribution. This led to Planck's intuitive curve-fitting approach, resulting in the elegant Planck's distribution law.

$$\rho(\nu, T)d\nu = \frac{2 \times 4\pi \nu^2 d\nu}{c^3} \frac{h\nu}{e^{\left(\frac{h\nu}{k_B T}\right)} - 1} \tag{1}$$

$\rho$ = energy density
T = temperature
$\nu$ = frequency
k= Boltzmann constant
c = speed of light
h = 6.627× $10^{-34}$ Joule-sec

Planck's derivation required the quantisation of the light modes, in equilibrium with the light-emitting matter of the blackbody cavity. The Planck's distribution $\frac{1}{e^{\left(\frac{h\nu}{k_B T}\right)} - 1}$ shows the probability of occupation of the energy state E = h$\nu$ at a temperature T. In the high-temperature case, it leads to the Boltzmann distribution function

(2)

$$P(E) = e^{-\frac{h\nu}{k_B T}}$$

Boltzmann's distribution can be derived by a counting argument in the context of canonical ensembles by distributing particles on the phase space shell [1].

It is worth mentioning that the phase space approach in classical mechanics, that treats coordinates and momenta on similar footing, was key to the success of statistical mechanics, originating from the phenomenological approach of thermodynamics.

Bose's venture into the derivation of Planck's distribution law started with the prodding of Saha, or probably by the classroom questions at Dhaka University, where he was employed as a Reader and his derivation through the counting of photon occupancy in phase space shells, brought in revolutionary ideas leading to the birth of Bose statistics, that governs the particle world of Bosons. The remaining particles known as Fermions, are described by the Fermi-Dirac statistics.

## 2. History of Bose statistics

### 2.1 Black-body radiation law, Planck's approach:

To understand the novelty and importance of Bose's work, it's essential to first review Planck's original derivation of the black-body radiation law.

$$\rho(\nu, T) = \frac{2 \times 4\pi h \nu^3}{c^3} \frac{1}{e^{\left(\frac{h\nu}{K_B T}\right)} - 1} \tag{3}$$

Planck's derivation involved three key steps:

1. Based on 'Classical Electromagnetic Theory': Planck established a relationship between the energy density ($\rho_\nu$) of incident radiation at temperature T with frequency between $\nu$ to $(\nu + d\nu)$, and

$$\rho_\nu = \frac{8\pi \nu^2}{c^3} U_\nu \tag{4}$$

Comparing above equations, he found the value of $U_\nu$
The average energy $U_\nu$ of a resonator at the same frequency and temperature.

$$U_\nu = \frac{h\nu}{e^{h\nu/K_B T} - 1} \quad (5)$$

2. In the second step, Planck calculated the entropy of oscillators by integrating the equation Tds = dU, where T is a function of U.

$$S = K_B \left[\left(1 + \frac{U_\nu}{h\nu}\right) ln\left(1 + \frac{U_\nu}{h\nu}\right) - \frac{U_\nu}{h\nu} ln \frac{U_\nu}{h\nu}\right] \quad (6)$$

3. In a revolutionary step, Planck introduced two ideas: He assumed that the total energy $U_N = NU_\nu$ of N oscillators was composed of discrete energy elements $\epsilon$, such that $U_N = P\epsilon$ (where P is a large number). He used Boltzmann's combinatorial approach, searching for a measure $W_N$ (the total number of distributions of energy values) that would correspond to his entropy equation.

$$S_N = k ln W_N \quad (7)$$

By using Stirling's formula and certain assumptions, Planck arrived at his quantum formula $\epsilon = h\nu$, which marked the birth of quantum theory. He admitted, however, that his combinatorial approach differed from Boltzmann's probabilistic method, leading to a major shift from classical physics. Without his bold, non-classical assumptions, Planck would have arrived at the classical Rayleigh-Jeans law. This derivation paved the way for future quantum developments, including the ground-breaking work of Bose. [2]

**2.2 Einstein's theory of light-quantum :**

In 1905, Einstein argued that the radiation following Wien's law exhibited energy fluctuations similar to those of material particles, suggesting that radiation must consist of discrete light-quanta, of magnitude

$$\left(\frac{R}{N_0}\right)\beta\nu = h\nu. \tag{8}$$

He applied this light-quantum hypothesis to explain phenomena such as the photoelectric effect. These findings, however, contradicted the classical electromagnetic theory that Planck had used to develop his theory of heat radiation. Einstein considered Planck's theory to be a counterpart to his own and critically analysed it. He concluded that Planck's quantum theory of radiation was based on the principle that the energy of a resonator could only take on discrete values, which are integral multiples of $h\nu$, and that energy changes occurred in jumps through absorption or emission. This finding contradicted classical electromagnetic theory, which did not account for such discrete energy values. Two years later, Lorentz reached the same conclusion. Although Planck initially tried to reconcile his theory with classical electrodynamics, it became clear that black-body radiation could not be fully explained using Maxwell's classical electrodynamics and the statistical mechanics of Maxwell and Boltzmann [2].

**2.3 Debye's explanation of Planck's law:**

In 1910, Debye offered a new derivation of Planck's radiation law to address the inconsistencies pointed out by Einstein regarding Planck's earlier work. Rather than using the relationship between radiation density and oscillator energy, Debye calculated the probability of a given radiation state and its entropy, bypassing the use of resonators. Following Rayleigh and Jeans, he determined that the number of elementary states or vibrational modes ($N_\nu d\nu$) within a volume (V) is:

$$N_\nu d\nu = \frac{8\pi\nu^2 V d\nu}{c^3} \tag{9}$$

He then assumed that the amount of energy $h\nu$ was distributed among these modes based on an arbitrary distribution function $f_\nu$, he obtained

$$\rho_\nu d\nu = \frac{8\pi h\nu^3}{c^3} f_\nu d\nu \tag{10}$$

Debye, similar to Planck, applied statistical methods to distribute energy quanta among the vibrational modes. He calculated the distribution function by maximising the number of possible configurations, subject to the constraint of constant energy [2].

From this, he determined the equilibrium entropy and derived Planck's radiation formula.

$$f_\nu = \frac{1}{e^{h\nu/K_B t} - 1} \tag{11}$$

Two key aspects of Debye's derivation, which influenced Bose's work fourteen years later, are important:

i. Debye showed that Planck's law arises solely from the assumption that energy transfer between matter and radiation is quantized in units of $h\nu$, without considering material resonators' properties. This aligned with Planck's views, but not Einstein's.

ii. Debye used Planck's definition of the probabilities $\omega_\nu$ without analysing their deeper significance.

## 2.4 Planck's quanta and the concept of identical indistinguishable particles:

In 1911, Natanson critically analysed the Planck-Debye combinatorial method and demonstrated that it involved distributing indistinguishable energy elements $\epsilon$ among N "receptacles of energy" distinguished by the numbers j quanta, contrary to the common belief that quanta were considered distinguishable. This insight, later supported by Ehrenfest and Kamerlingh Onnes in 1914, revealed that Planck and Debye had implicitly assumed the indistinguishability of quanta. However, Ehrenfest and Onnes argued that Planck's method distributed indistinguishable energy elements among distinguishable resonators, contrasting with Einstein's concept of statistically independent light quanta. They concluded that Planck's approach could not be interpreted in the same way as Einstein's light-quanta theory [2].

**2.5 The concept of phase-space cells by Planck:**

Planck realized that classical statistical mechanics had to be modified to derive his radiation formula, rather than the Rayleigh-Jeans law. He shared this conclusion at the 1911 Solvay Congress. According to Gibbs, the probability of finding a single particle in the element $d^3p\, d^3q$ of the six-dimensional phase space is given by $\dfrac{e^{\frac{-E}{K_BT}}d^3p\,d^3q}{\int e^{\frac{-E}{K_BT}}d^3p\,d^3q}$ .

According to classical mechanics, the average energy of a one-dimensional oscillator is proportional to temperature. However, if energy is quantized as integral multiples of $h\nu$. Planck derived that the energy of an oscillator is discrete.

$$U = \frac{\sum_n E_n\, e^{-E_n/K_BT}}{\sum_n e^{-E_n/K_BT}} = \frac{\epsilon}{e^{\epsilon/K_BT} - 1} \qquad (12)$$

This led him to interpret the constant h as defining the finite extension of the elementary areas in phase space, meaning that oscillators' energies must be quantised.

Planck concluded that energy quanta arose from a fundamental condition, requiring a revision of classical mechanics, which could not explain this quantum behaviour. He declared that all attempts to reconcile classical mechanics with quantum theory should be abandoned. These ideas deeply influenced Bose, who recognised the need to modify classical electrodynamics and statistical mechanics. Bose saw that quantum states are not continuously distributed in phase space, allowing them to be counted by dividing the total phase space volume by $h^3$, and accepted Planck's argument that classical mechanics needed adjustments to align with quantum theory.

**2.6 Spontaneous and induced transitions:**

In 1917, Einstein made significant progress in explaining Planck's radiation law by using Bohr's model of atoms with discrete energy states. He derived Planck's law by assuming that energy transitions between atomic states occurred through the exchange of energy quanta $h\nu$ without relying on classical electrodynamics. He introduced the concepts of spontaneous emission (where atoms emit radiation independently) and induced emission (where radiation is triggered by an external field). Using principles like Boltzmann's statistical theory he wrote the probability $W_n$ for an atom to be in a stationary state with quantum number n and $g_n$ (statistical weight of the state) in the form:

$$W_n = g_n exp\left(-\frac{\epsilon_n}{K_B T}\right) \tag{13}$$

Einstein recognized that while his hypotheses for radiation interaction were elegant, they were not fully proven. By 1916, the light-quantum hypothesis had gained empirical support from Millikan's verification of Einstein's photoelectric equation. However, Einstein acknowledged that his theory did not yet reconcile with wave theory and left certain processes to chance, which left him unsatisfied. Despite this, Einstein's work laid the foundation for future quantum theory, influencing Bose's later contributions, though Bose did not adopt all of Einstein's assumptions.

**2.7 Discovery of Compton effect:**

Most physicists initially rejected Einstein's light-quantum hypothesis, even after Millikan confirmed the photoelectric equation in 1916. Millikan himself criticised the semi-corpuscular theory behind it. The main challenge was reconciling light-quanta with interference and diffraction phenomena, which strongly supported the wave theory of light. In 1912, the wave nature of X-rays was confirmed by Max von Laue's experiments, reinforcing scepticism toward the light-quanta idea. Even though Planck, Nernst, and others recognised Einstein's brilliance, they viewed the light-quantum hypothesis with scepticism.
Bohr, too, resisted the hypothesis, as it clashed with his aim to build atomic theory on the correspondence principle. However, the hypothesis gained acceptance in 1923 when Compton and Debye independently demonstrated that the scattering of X-rays by atoms, observed by Compton, could only be explained by treating X-rays as particles. This discovery showed that energy and momentum were conserved in these scattering processes, providing strong evidence for the quantum nature of light. Arnold Sommerfeld, upon learning of Compton's work, advocated for its importance in quantum theory, discussing it with prominent physicists like Einstein, leading to wider recognition of Compton's findings.

**2.8 Pauli's formulation of Planck's blackbody radiation law:**

In 1923, Pauli critically examined the equilibrium of radiation with electrons. He concluded that for electrons having a Maxwellian distribution of velocity and radiation density by Planck's law are made to consider reversible collision taken in to account Energy momentum conservation under the assumption that the collision probability for unit time is given by, $A\rho_\nu + B\rho_\nu$ where A and B are the coefficients of Einstein's theory of radiation and $\rho_\nu$, are the radiation densities before and after the collision respectively. Pauli found that different terms dominated for different radiation regimes: Wien's law for short wavelengths and Rayleigh-Jeans law for longer wavelengths.

Einstein and Ehrenfest praised Pauli's work, expanding it to involve interactions with more than two light-quanta. They also addressed a paradoxical aspect of Pauli's findings, where the number of scattering events per unit time increased faster than proportional to the radiation density after the collision. They concluded that this was due to the inclusion of induced emission processes (Einstein's "negative radiation").

## 3. Bose on light quanta:

In 1924, S.N. Bose, while at Dhaka University, re-derived Planck's radiation law by deriving a new form of statistics specific to light quanta, which later became known as 'Bose-Einstein statistics'. He sent his manuscript to Einstein and after the translation of it into German he published it in *Zeitschrift für Physik*, praising it as a significant advancement. However, Einstein made a notable change to Bose's original manuscript. Bose proposed that every photon has helicity (intrinsic angular momentum ) of one quantum unit, either parallel or antiparallel to its direction of motion. Einstein found this idea too radical and removed it from the final publication.

Later, experimental confirmation of this intrinsic angular momentum (spin) came in 1931 through a paper by Raman and Bhagavantam. They verified that Bose's suggestion of a "spin" for the photon, requiring a factor of 2 for right-handed or left-handed spin states, was correct [3]. This factor explained the proper arrangement of quanta in phase space.

Bose's insight into the photon's angular momentum preceded the postulation of the electron's spin in 1925 by Uhlenbeck and Goudsmit. The term "photon" itself wasn't coined until 1926, and further experimental evidence, such as Richard Beth's work confirming photon angular momentum, came much later.
Bose likely deduced the photon's angular momentum by incorporating Einstein's energy equation into an earlier expression by Poynting, who had proposed a similar idea for circularly polarised electromagnetic waves in 1909. Bose's prediction of only two possible photon spin states ($\pm \hbar/2\pi$) was later confirmed by Eugene Wigner in 1939, using quantum field theory to show that massless particles like photons can only have two helicity states.

## 4. Scientific papers of Bose:

Though prominent European and American physicists were initially sceptical or dismissive of Einstein's theory on light quantum, two scientist from India, M. N. Saha and S. N. Bose, recognised its significance and applied it successfully in their work. After Einstein's 1917 paper, which suggested that light quanta carry directed momentum, Saha utilised this concept to explore radiation pressure on molecular-scale objects. Classical models by Nicholson and Klotz had previously argued that radiation pressure would have a negligible impact on such particles [4] . In contrast, Saha adopted the quantum theory of light, positing that light exists in discrete energy pulses. When a molecule absorbs one of these pulses, it gains momentum, moving forward. Saha concluded that radiation pressure could exert a substantial effect on atoms and molecules, well beyond what classical theories predicted—demonstrating the first practical use of Einstein's light-quantum hypothesis.

In early 1924, Saha visited Dhaka, where he stayed with Bose, who was then teaching postgraduate students and was wrestling with issues in Planck's law. Saha developed Bose's interest in to Pauli's 1923 work, which connected to Einstein and Ehrenfest's and Einstein's 1917 publications. Recalling this interaction, Bose noted that Pauli's insights suggested that applying quantum conditions required precise understanding of process outcomes.

Inspired by his discussions with Saha, Bose deeply examined the work of Planck, Debye, Einstein, Compton, Pauli, and Einstein and Ehrenfest. His research led to two important papers written in June 1924, which he forwarded to Einstein for review and publication. Einstein translated them into German, added comments, and arranged for their publication in 'Zeitschrift für Physik' later that year.

## 4.1 First paper of Bose (Planck'S law and the light quantum hypothesis)

Planck's formula for blackbody radiation, which laid the foundation for quantum theory, has been the subject of numerous derivations since its publication in 1901. However, all these derivations face a fundamental problem: The foundational assumptions of quantum theory are not fully aligned with the principles of classical electrodynamics. Existing derivations rely on a relationship between radiation density and the mean energy of an oscillator, using assumptions from classical theory about the ether's degrees of freedom, which leads to logical inconsistencies [5].

$$\rho_\nu d\nu = (8\pi \nu^2 d\nu / c^3) E \qquad (14)$$

Einstein provided a notably elegant derivation, aiming to avoid classical assumptions by focusing on simple energy exchange principles between molecules and radiation. However, to align his formula with Planck's, he had to use Wien's displacement law and Bohr's correspondence principle, both of which are rooted in classical theory. This reveals the limitations of these derivations, as they rely on classical concepts in some form.

Bose suggested that combining the light quantum hypothesis with statistical mechanics, as advocated by Planck for quantum theory, is a sufficient and independent method for deriving the law without needing classical theory. He proposed a new approach, involving radiation enclosed in a volume, where different types of quanta are characterized by their number and energy. The solution to the problem involves determining the distribution of quanta that maximizes the probability, subject to certain conditions.

$$E = \sum_s N_s h \nu_s = V \int \rho_\nu d\nu \qquad (15)$$

The quantum is characterized by its momentum, which has a magnitude of $\frac{h\nu_s}{c}$ (s = 0 to = ∞) in the direction of motion, along with its spatial coordinates and associated momenta. These six variables (coordinates and momenta) can be understood as points in a six-dimensional phase space. The frequency defines a cylindrical surface within this space, and the frequency range $d\nu_s$ corresponds to a particular volume of the phase space.

$$\int dx\, dy\, dz\, dp_x dp_y dp_z = \frac{V 4\pi \left(\frac{h\nu}{c}\right)^2 h d\nu}{c} = 4\pi(h^3\nu^2/c^3)V d\nu \qquad (16)$$

When the phase space is divided into cells of size $h^3$ the number of cells within the frequency domain $d\nu$ is given by $4\pi V(\nu^2/c^3)$ To account for polarization, this number is multiplied by 2, resulting in $8\pi V(\nu^2 d\nu/c^3)$.
This allows for the calculation of the thermodynamic probability of a macroscopically defined state. For a given frequency domain $d\nu_s$, the number of quanta $N_s$ can be distributed across the available cells in various ways. The number of possible distributions is determined by how the quanta are placed into vacant cells, with different possibilities for cells containing one, two, or more quanta. After some fundamental calculation he derived,

$$E = \sum_s \frac{8\pi h \nu_s^3}{c^3} V \left[ exp\left(\frac{h\nu_s}{kT}\right) - 1 \right]^{-1} d\nu_s \qquad (17)$$

Which is identical to Planck's formula.

### 4.2 Second paper of Bose (Thermal equilibrium In radiation Field In The presence of matter):

In his paper completed on June 14, 1924, Bose presented two main contributions. He established general conditions for the statistical equilibrium of a system composed of matter and radiation, without making any specific assumptions regarding the mechanisms of elementary radiative processes. Additionally, he introduced a new expression for the probability of these processes, which deviated from Einstein's formulation. However, Einstein critiqued Bose's approach, stating it was not applicable to elementary radiative processes, which led to the paper being largely ignored. Despite this, Bose maintained that Einstein had not fully appreciated his work.

Bose began by critically reviewing key derivations of Planck's law by Debye (1910), Einstein (1917), Pauli (1923), and Einstein and Ehrenfest (1923). He argued that Debye's derivation still relied on classical electrodynamics and that the other derivations were based on ad hoc assumptions about the probabilities of radiative processes. Bose believed that the problem of thermodynamic equilibrium of radiation in the presence of material particles could be studied using statistical mechanics, independently of these assumptions. He proposed a general relation that would be valid regardless of the specifics of elementary processes, in line with Kirchhoff's law.

Bose further argued that the thermodynamic probability for the combined system of matter and radiation could be obtained by multiplying the probabilities of radiation and energy distribution among particles. The equilibrium condition would then be that this combined probability is maximized. For the radiation's thermodynamic probability, Bose introduced a specific expression [2].

$$W = \prod_s \frac{(A_s + N_s dv)!}{A_s! N_s dv!} \qquad (18)$$

Where $A_s = \frac{8\pi V v^2 dv}{c^3}$, attributing it to both his earlier work and Debye's 1910 paper. Notably, while Bose cites his own previous paper as appearing in *Philosophical Magazine*, this was due to the fact that he initially submitted it there before later sending it to Einstein. During translation, the reference was not updated to reflect its actual publication in 'Zeitschrift für Physik'

Although Debye's Ansatz is referenced, it does not appear in Bose's first paper. Instead, an equivalent expression from Natanson appears. Bose did not clarify this equivalence, likely assuming it was already well known in the literature of the time.

For the thermodynamic probability of material particles, Bose extends his assumptions to include cases like the Bohr atom with discrete energy levels, as well as particles' translational energy. He divides the phase space into cells, assigning a probability 'g' for a particle to occupy a cell. These probabilities are generally equal, except for the Bohr atom. The resulting thermodynamic probability follows the classical Maxwell-Boltzmann distribution for material particles, treated as distinguishable, while light quanta are treated using quantum mechanics, where they are indistinguishable. He then considers an elementary

$$W = \prod_s \frac{(A_s + N_s)!}{A_s! N_s!} \prod_r \frac{g_r + N!}{n_r!} \quad (19)$$

$$\sum N_s h v_s = E \quad (20)$$

process in which a particle passes from the rth cell to the sth cell while a light-quantum of frequency $v$ changes into a light-quantum of frequency $v'$. The stationarity of W gives

$$\frac{n_r}{g_r} \prod_v \frac{N_v}{N_v + A_v} = \frac{N_s}{g_s} \prod_{v'} \frac{N_{v'}}{N_{v'} + A_{v'}} \quad (21)$$

Where $\sum h v' - \sum h v + E_s - E_r = 0$

Bose subsequently showed that he successfully extended the results of Pauli (1923) and Einstein and Ehrenfest (1923) without requiring any arbitrary assumptions about elementary radiative processes to derive Planck's formula. Specifically, this applied to the case of Bohr's atoms, which Einstein had examined in 1917.

$$\frac{n_r}{g_r} \frac{N_v}{N_v + A_v} = \frac{n_s}{g_s} \quad (22)$$

Einstein's derivation of conditions for atomic transitions, which occur in two ways:
1. 'Spontaneous transitions', which are independent of the external radiation field (similar to radioactivity).

2. 'Induced transitions', whose probability depends on the external radiation field. These occur when atoms move from lower to higher energy levels via induced absorption, also dependent on the radiation field.

Einstein had to assume certain relationships between transition probabilities, particularly Wien's and Rayleigh-Jeans laws, to derive Planck's law. In contrast, Bose's derivation was independent of these assumptions, prompting Einstein to translate and publish Bose's work. However, their disagreement began with Bose's new proposed probability for the interaction between particles and radiation quanta.

Bose argued that in a collision, no interaction is as probable as any specific interaction, which is a departure from classical theory. This idea parallels Bohr's theory of stationary states, suggesting that, similar to electron-atom collisions, particles may pass through radiation without interaction. Bose refers to the Ramsauer-Townsend effect, where electrons pass through atoms without altering their motion.

He introduced the idea of "cells" in phase space, where the probability of interaction between radiation and particles depends on the number of quanta in a cell. When a particle and quanta occupy the same cell, there are (r + 1) possible outcomes, including various levels of energy exchange or no interaction at all. The number of cases in which interaction or energy exchange occurs is

$$p_1 + 2p_2 + 3p_3 + \cdots = N_s dv_s = \sum_r r p_r \qquad (23)$$

Consequently the probability of an interaction is

$$p = \frac{\sum r p_r}{\sum (r+1) p_r} = \frac{N_s dv_s}{A_s + N_v dv} \qquad (24)$$

This is Bose's second fundamental result

Einstein ensured that both of Bose's papers were published in *'Zeitschrift für Physik'*, yet his responses to each differed notably. Einstein was highly supportive of Bose's first paper, seeing it as an important contribution. However, his extensive comments on the second paper took a more critical tone, noting two issues:

*"Your principle is not compatible with the following two conditions:
(1) The absorption coefficient is independent of the radiation density.
(2) The behaviour of a resonator in a radiation field should follow from the statistical laws as a limiting case."*

Bose was disappointed by these comments and, on January 27, 1925, he sent Einstein a rebuttal, saying:

"I have written down my ideas in the form of a paper [his third] which I send under separate cover … I have tried to look at the radiation field from a new standpoint and have sought to separate the propagation of quantum of energy from the propagation of electromagnetic influence…"

Bose later claimed that he had provided a quantum mechanical explanation for the factor of two in his paper. However, Einstein reportedly removed this and replaced it with an argument based on polarization. The original English manuscript of Bose's paper is missing from the archives, but it is believed that Bose had suggested that light-quanta possess an intrinsic spin with values of $\pm h/2\pi$.

## 5. Bose-Einstein condensate:

A modern explanation describes that, at elevated temperatures, particles in a Bose gas are spread across various energy levels as determined by Bose-Einstein statistics. However, when the temperature drops below a specific critical point, a large portion of particles condense into the lowest energy state, creating what is known as a Bose-Einstein Condensate (BEC). This state allows quantum effects to be observed on a macroscopic scale, providing a unique opportunity to study fundamental properties of matter.

## 6. Conclusion:

Satyendra Nath Bose's contributions to scientific research and education are widely celebrated. Appointed President of the Indian Science Congress in 1945, he held this role until 1948, significantly advancing Indian scientific dialogue. Concurrently, he served as President of the Indian National Science Academy from 1948 to 1950. In recognition of his achievements, Bose was elected as a Fellow of the Royal Society in 1958. He passed away on February 4, 1974, at the age of 80. Honouring his legacy, the Government of India founded the "Satyendra Nath Bose National Centre for Basic Sciences" in Kolkata in 1986. His dedication to fundamental science and his lasting impact on Indian science remain celebrated to this day.

## Acknowledgement:

We would like to acknowledge Prof. Bedanga Mohanty for organising the insightful workshop on BOSESTAT@100 at NISER, and Prof. Lambodar Prasad Singh for arranging a similar talk at Udayanath Autonomous College of science and technology, Adaspur, which provided the impetus for this article.